\documentclass[a4paper]{llncs}

\usepackage{amssymb}
\usepackage{graphicx}
\usepackage{algpseudocode}
\usepackage{algorithm}
\usepackage{hyperref} 
\usepackage[utf8]{inputenc}
\usepackage[english]{babel}

\usepackage{csquotes}
\MakeOuterQuote{"}

\usepackage{amsmath}

\usepackage{graphicx}
\usepackage{algpseudocode}
\usepackage{algorithm}
\usepackage[utf8]{inputenc}
\usepackage[english]{babel}
\usepackage{csquotes}
\usepackage{mathrsfs}
\usepackage{yfonts}
\usepackage{amsmath}
\usepackage{tikz}
\usetikzlibrary{calc}
\usepackage{relsize}
\usepackage{xfrac}

\usepackage{lipsum}
\usepackage{graphicx}

\usepackage{varwidth}
\usepackage{afterpage}
\usepackage{lipsum}
\usepackage{amssymb}

\begin{document}

\mainmatter  

\title{RDV: An Alternative To Proof-of-Work\\
	And A Real Decentralized Consensus For Blockchain}

\titlerunning{RDV: An Alternative To Proof-of-Work}

%
%
\author{Siamak Solat}
\authorrunning{RDV}

\institute{}

%
%

\toctitle{: An Alternative To Proof-of-Work}
\tocauthor{Authors' Instructions}
\maketitle

\begin{abstract}
	
A blockchain is a decentralized ledger where all transactions are recorded. To achieve immutability of transactions history, we need a real decentralized consensus and permission-less blockchain since in a permissioned blockchain, although we can accelerate transactions validation throughput, however contrary to permission-less blockchains that are open to everybody for participating in transactions validation process, in a permissioned blockchain the fate of transactions is controlled by a limited number of validators such that this fact can impair decentralization of the system. Bitcoin as a permission-less blockchain uses proof-of-work (PoW). PoW powered blockchains currently account for more than 90\% of the total market capitalization of existing digital currencies \cite{PoW_blockchains}. PoW is a cryptographic puzzle that is difficult to solve but easy to verify. However, significant latency of proof-of-work for transactions confirmation has negative effects on blockchain security such that longer delays may increase the number of forks and the possibilities for mounting double-spending attacks \cite{misbehavior_in_bitcoin}. On the other hand, PoW consumes a significant amount of energy that by growing the network, it becomes a major problematic of this consensus mechanism. We introduce an alternative to PoW, because of all its major problems and security issues that may lead to collapsing decentralization of the blockchain, while a full decentralized system is the main purpose of using blockchain technology. The approach we introduce is based on a distributed voting process and called “RDV: Register, Deposit, Vote” in which all participants by proceeding a registration step can participate in voting process in a permission-less blockchain. Since in RDV algorithm, there is no mining process, so it may be more appropriate for low-level energy devices and Internet of Things (IoT).


\end{abstract}

\section{Introduction}

Bitcoin blockchain \cite{Nakamoto} was introduced as a peer-to-peer system aims at fully decentralization of transactions. 
For this purpose, we need a \textbf{reliable} and \textbf{immutable} blockcahin. It is reachable by a secure and \textbf{decentralized} consensus mechanism. One of the well-known consensus approaches is proof-of-work. It is a cryptographic puzzle that is difficult to solve but easy to verify. A considerable latency in inter-block time increases possibility of double-spending attack \cite{misbehavior_in_bitcoin}. On the other side, a miner who controls a significant number of nodes in a mining pool is able to increase the winning probability of their branch in a fork by passing their own blocks and rejecting the others \cite{Sok}. When a temporary fork occurs, it is possible for an attacker to make a double-spending attack \cite{Bamert} \cite{Double-spending}. 
Despite belief that PoW has an acceptable scalability as a  lottery-based algorithm due to no need to exchange messages \cite{Hyperledger_Consensus} ; however, if we define PoW as one-cpu-one-vote, then with growing the network hashing power of the network will be increased and as a result, we need to increase difficulty of PoW that causes participating in transactions validation (aka mining process) would be more difficult for miners who do not possess enough fast processors (CPU / GPU / ASIC etc) and this situation continues till for participating in mining process must be joined to a large mining pool. This process, in a long time, causes the blockchain would be controlled by some large and limited mining pools. This eventually affects negatively decentralization of the blockchain.
On the other hand, PoW consumes a significant amount of energy that by growing the network, it becomes a major problematic of this consensus mechanism. This causes also difficulties in order to use blockchain for low-level energy devices and IoT. In RDV algorithm, there is no mining process and it does not consume a considerable energy. This feature causes also the RDV consensus mechanism would be appropriate for low-level energy devices and Internet of Things.

\subsection{Definitions}
\label{Definitions}

\begin{definition}
	\normalfont
	\textit{voting process}: \par
		Every transaction for being inserted in a block needs to be participated in a voting process in which if majority of current voters in the \textit{voteRbox} vote for this transaction, it will be inserted in a new block in the blockchain.     	
\end{definition}

\begin{definition}
	\normalfont
	\textit{registered node}: \par 
	Every node for participating in a voting process needs to register by pledging a part of their coins, meaning that if the amount needed for registration is \textit{d} coins and the balance of candidate node is \textit{c} coins, then after registration their balance becomes \textit{(c - d)} coins. 
\end{definition}

\begin{definition}
	\normalfont
	\textit{ordinary node}: \par 
	Once the voter node decides to leave registration mode to participate in the network only as an \textit{ordinary} node, then this part of his coins (i.e. \textit{d} coins) will be unblocked such as the node will be able to use it again. In this mode, the node is not permitted to participate in voting process.
\end{definition}

\begin{definition}
	\normalfont
	\textit{time} $ \Delta $: \par 
	We consider the time $\Delta$ because we prefer to have the vote of all voters to achieve better decentralized voting process and on the other hand, maybe some voters do not participate in some voting processes. So, we choose a reasonable duration for the time $\Delta$, such that a registered node must participate at least in a voting process in every $\Delta$ time unit if at least a transaction has been sent to the network since $\Delta$ time unit. The duration of the time $\Delta$ depends on incoming transaction rate in the network. If rate of incoming transactions is high, then we choose a smaller value for $\Delta$ and vice versa. Because in case of arriving more transactions, we expect that voters participate in more voting process. Also, information propagation delay time affects duration of $\Delta$ time. For example, information propagation time has been calculated for the Bitcoin network by authors in \cite{Decker}, so this delay time is calculable for any other similar network. Eventually, according to this delay time and incoming transaction rate we can calculate duration of $\Delta$ time.
\end{definition}

\begin{definition}
	\normalfont
	\textit{time} $ \Pi $: \par 
	If there is a voter who has not participated in any voting process for $\Delta$ time unit while a transaction is sent since $\Delta$ time unit, their identity is removed from \textit{voteRbox} for $\Pi$ time unit.
	We choose duration of the time $\Pi$ neither very long such that nodes lose their motivation to continue as a voter nor very short such that nodes do not sense a penalty. We consider this penalty in order that if nodes register as a voter, then they have to participate in voting process as much as possible, otherwise if they do not intend to vote, they must leave registration mode to become an ordinary node.
\end{definition}

\begin{definition}
	\normalfont
	\textit{Priority Point}: \par 
	In RDV, the nodes do not decide which transaction must be participated in voting process, but also there is a parameter, \textbf{Priority Point}, by which a transaction among others will be selected to be participated in voting process. This parameter is calculated as follows:
	
	\begin{equation} 
	\label{eq-Priority-Point}
	\mathrm{(tx \rightarrow prp) = [curTi - (tx \rightarrow tsp)] + (tx \rightarrow CTR)}
	\end{equation}
	
	Where, $ \mathrm{(tx \rightarrow prp)} $ is the priority point of transaction, $ \mathrm{(tx \rightarrow tsp)} $ is the time at which transaction has been sent, curTi is the current time and $ \mathrm{(tx \rightarrow CTR)} $ is a "Confirmation Time Reward" (as defined in below). Then, transaction with most Priority Point will be selected to be participated in voting process (see Table \ref{TablePeriority}).
\end{definition}

\begin{definition}
	\normalfont
	\textit{CTR Parameter (Confirmation Time Reward)}: \par 
	If the result of voting process equals to vote of a voter, then a unit will be added to CTR parameter of transactions belong to this voter. So, CTR parameter helps voters to increase the Priority Point of their transactions as an incentive. As a result, CTR parameter incentivizes voters to participate \textbf{as much as possible} in voting processes and validating \textbf{correctly} transactions. Thus, we do not need any resource such as transactions fee to provide the monetary rewards, such that we can support fee-free transactions. Whereas, in a PoW-based system, it is crucial to \textbf{incentivize} miners by \textbf{monetary} reward, because mining process has \textbf{considerable monetary cost} (i.e. energy/electricity along with providing hardware cost) and so participating in mining process must be affordable and economic for miners. 
	However, in RDV, participating in voting process has not a considerable monetary/energy cost and so we can use other incentives such as "Confirmation Time Reward" (as defined above). \textbf{Although, it is possible to exchange this CTR rewards with coins between users}. For example, assume user A possesses some CTRs and user B has some coins. So they can exchange CTRs and coins between each other by a multi-signature transaction, meaning that they are exchangeable after both users sign the exchange transaction.
\end{definition}


\section{RDV Consensus Algorithm}

RDV algorithm has an incentive-punitive mechanism and is based on distributed voting process. It includes three main steps as follows: (1) Register (2) Deposit (3) Vote. We then describe the details of RDV algorithm in Section \ref{RDV_Flowchart}.

\begin{itemize} 
	\item \textit{Register: } 
	Every node to be authorized to vote for a transaction has to register. Otherwise, the node participates as an "ordinary node" which is only able to send transactions. All registered nodes are stored in the blockchain. 
	\item \textit{Deposit: }
	It means pledging some coins as collateral to be able to finalise successfully registration step. The necessary amount of coins for \textit{deposit} step is calculable regarding to the price of a coin. 
	The registration process is acceptable if and only if a part of the coins of the candidate node who intends to participate in voting process has been deposited, meaning that the registered nodes have no access to this part of their coins as long as the node is a "voter". For example, if the amount needed for registration is \textit{d} coins and the balance of candidate node is \textit{c} coins, then after registration, their balance becomes \textit{(c - d)} coins. Once the voter node decides to leave registration mode to participate in the network only as an \textit{ordinary} node, then this part of their coins (i.e. \textit{d} coins) will be unblocked such that the node will be able to use it again.
	\item \textit{Vote: } 
	Every registered node is permitted to vote for transactions, either positive (i.e. 1) or negative (i.e. 0). 
	Then, voter signs the vote. In case the result of voting process is not equal to a voter's vote, then this voter will lose a part of their deposited coins "for ever" as a penalty. We consider this penalty to prevent malicious behaviours. The amount of this penalty is also calculable (like \textit{Deposit} step). On the other side, if the result of voting process is equal to the voter's vote, then this voter receives CTR reward that is exchangeable with coins by a multi-signature transaction as described in Section \ref{Definitions}. 
\end{itemize}

\subsection{RDV Algorithm Flowchart}
\label{RDV_Flowchart}
\begin{figure*}
	\makebox[\textwidth][c]{\includegraphics[width=0.9\textwidth]{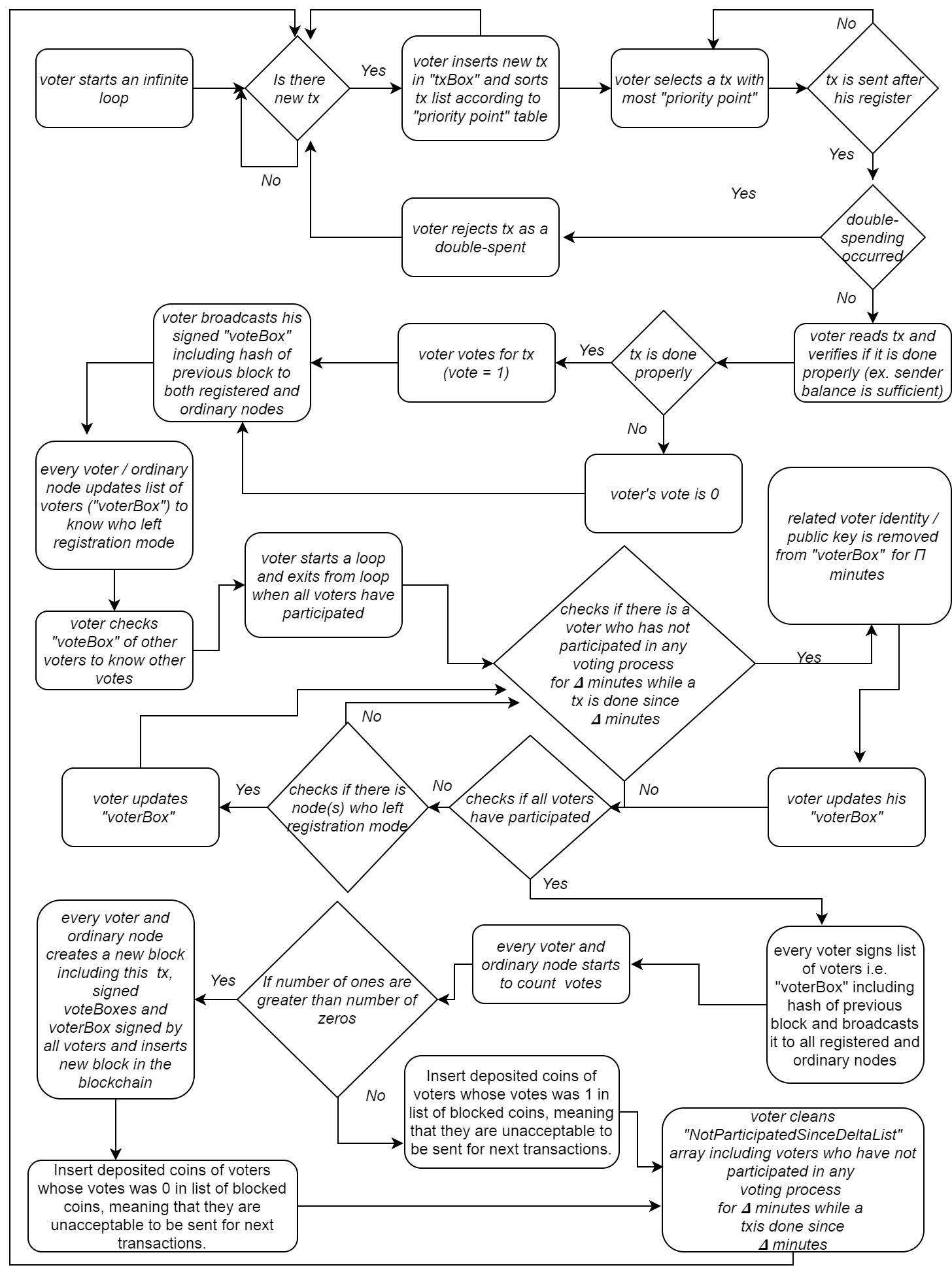}}
	\caption{RDV Algorithm Flowchart.}
	\label{fig_2}
\end{figure*}

In Figure \ref{fig_2} we show the circle of what a node does for every transaction. After registration, voter starts an infinite loop. Then voter checks if there is a new transaction. If so, voter inserts new \textit{tx} in his \textit{txBox} and sorts transactions by "Priority Point" table (see Table \ref{TablePeriority}) such that the first transaction in the list (i.e. txBox[0]) has more \textit{Priority Point} to be participated in voting process. Then voter selects a transaction with most \textit{Priority Point} (i.e. txBox[0]). The voter checks only transactions that have been sent after his registration, so voter checks if this \textit{tx} is sent after his registration. If so, voter checks if this \textit{tx} is double-spent. Double-spending is checked using Table \ref{TablePeriority} (see Section \ref{Double-Spending}). If \textit{tx} is double-spent, voter rejects this \textit{tx} as a double-spent and waits for another new transaction. Otherwise, voter checks if \textit{tx} is done properly (ex. sender balance is sufficient). If everything is fine, voter votes for \textit{tx} (i.e. vote = 1). If there is something wrong voter's vote would be 0. Afterwards, voter broadcasts his signed \texttt{voteBox[txID][voterID][hash of previous block]} including hash of previous block to both registered and ordinary nodes. 

Note that all registered and ordinary nodes have list of voters in \textit{voteRbox}. Then every voter and ordinary node updates list of voters ("\textit{voteRbox}") to know who left registration mode. voter then checks his \textit{voteBox} where they receive and keep vote of other voters. Afterwards, voter starts a loop and exits from loop when all voters have participated in voting process. 

Then voter checks if there is a voter who has not participated in any voting process for $\Delta$ time unit while a transaction is sent since $\Delta$ time unit. We consider the time $\Delta$ because we prefer to have the vote of all voters to achieve better decentralized voting process and on the other hand, maybe some voters do not participate in some voting processes. So, we choose a reasonable duration for the time $\Delta$, such that a registered node must participate at least in a voting process in every $\Delta$ time unit if at least a transaction has been sent to the network since $\Delta$ time unit. The duration of the time $\Delta$ depends on \textbf{incoming transaction rate} in the network. If rate of incoming transactions is high, then we choose a smaller value for $\Delta$ and vice versa. Because in case of arriving more transactions, we expect that voters participate in more voting process. Then if there is such a voter, their identity is removed from \textit{voteRbox} 
for $\Pi$ time unit. 
We choose duration of the time $\Pi$ neither very long such that nodes lose their motivation to continue as a voter nor very short such that nodes do not sense a penalty. We consider this penalty in order that if nodes register as a voter, then they have to participate in voting process as much as possible, otherwise if they do not intend to vote, they must leave registration mode to become an ordinary node. Because we need vote of all existing registered nodes in current\textit{voteRbox}. Afterwards, voter updates \textit{voteRbox} by removing such registered nodes. Then voter checks if all voters have participated in voting process. If not, voter checks if there is node(s) who left registration mode. If so, voter updates \textit{voteRbox}. The voter continue this loop till all registered nodes in current \textit{voteRbox} participate in voting process. Note that as we mentioned above, if a registered node does not participate in any voting process for $\Delta$ time unit while at least a transaction has been sent since $\Delta$ time unit, then related node will be removed from \textit{voteRbox} and so eventually we achieve a point at which all registered nodes in \textit{voteRbox} have participated in voting process (see Section \ref{Correctness}). Then every voter signs list of voters i.e. \textit{voteRbox} including hash of previous block and broadcasts it to all registered and ordinary nodes. Every voter and ordinary node then starts to count votes. If number of "1" are greater than number of "0", every voter and ordinary node creates a new block including this \textit{tx}, signed \textit{voteBoxes} and \textit{voteRbox} signed by all voters and inserts new block in the blockchain and puts deposited coins of voters whose votes was "0" in list of "blocked coins", meaning that they are unacceptable to be sent for next transactions by those nodes. If number of "0" are greater than number of "1", every voter and ordinary node puts deposited coins of voters whose votes was "1" in list of "blocked coins" to be unacceptable for the next transactions. Finally, voter cleans \textit{NotParticipate} array includes registered nodes that have not participated in any voting process for $\Delta$ time unit while at least a \textit{tx} has been sent since $ \Delta $ time unit. Then, voter waits for receiving another new transaction. 

\afterpage{ 
	\clearpage
	\afterpage{%
		\thispagestyle{empty}
	}
	\begin{algorithm*}
		\caption{RDV Algorithm - Voting Process}\label{rdv}
		\begin{algorithmic}[1]
			
			\Require
			\Statex txBox[tx list] \Comment{keeps list of transactions a voter receives from the network.}
			
			\Statex voteBox[tx.ID][voterID][hash of previous block] \Comment{keeps votes value.}
			
			\Statex voteRbox[tx.ID][list of voters][hash of previous block] \Comment{keeps list of voters ID.}

			\While{()} \label{rdvline:1}
			
			\State tx $ \gets $ isThereNeWtx()  \label{rdvline:2}
			
			\If{(tx $ \neq $ null)} \label{rdvline:3}	     
			\State txBox[tx list] $ \gets $ SorTtxList(tx.time-stamp , txBox[tx list]) \label{rdvline:4}	    
			\EndIf \label{rdvline:5}	   
			
			\State tx $ \gets $ txBox[0]  \label{rdvline:6}
			
			\If{(tx.time-stamp $ > $ voterRegisterTime)} \label{rdvline:7}
			\If{(Result of checking for douple-spending = true)} \label{rdvline:8}
			\State remark tx as a double-spent and goto line 2. \label{rdvline:9}
			\ElsIf {(verify(tx) = true)} \label{rdvline:10}
			\State Sign(voteBox[tx.ID][voterID][hash of previous block] $ \gets $ 1) \label{rdvline:11}
			\Else \label{rdvline:12}
			\State Sign(voteBox[tx.ID][voterID][hash of previous block] $ \gets $ 0)  \label{rdvline:13}
			\EndIf \label{rdvline:14}
			
			\State broadcast signed voteBox  \label{rdvline:15}
			
			\State update voteRbox to know who left registration mode  \label{rdvline:16}
			
			\State check voteBox of other voters to know other votes \label{rdvline:17}
			
			\While{(AllVoterHaveParticipated(voteRbox)$ \neq $ true)} \label{rdvline:18}
			
			\State NotParticipatedSinceDeltaList $ \gets $ NotParticipatedSinceDelta() \label{rdvline:19}
			
			\If{(NotParticipatedSinceDeltaList $ \neq $ null)} \label{rdvline:20}
			\State Remove voters exist in this list from voteRbox for $ \Pi $ time unit. \label{rdvline:21} 
			\State update voteRbox \label{rdvline:22}
			\EndIf \label{rdvline:23}
			
			\If{(AllVoterHaveParticipated(voteRbox)$ \neq $ true)} \label{rdvline:24}
			\If{(there is node(s) who left registration mode)} \label{rdvline:25}
			\State update "voteRbox"  \label{rdvline:26}
			\EndIf \label{rdvline:27}
			\EndIf \label{rdvline:28}
			
			\EndWhile \label{rdvline:29}
			
			\State Sign(voteRbox[tx.ID][list of voters][hash of previous block]) \label{rdvline:30}
			
			\State broadcast Sign(voteRbox) \label{rdvline:31}
			
			\State Start to count votes  \label{rdvline:32}
			
			\If{(tx.ID.Ones $>$ tx.ID.Zeros} \label{rdvline:33}
			\State Create block includes voteBoxes and voteRbox signed by all voters \label{rdvline:34}		
			\State Insert new block in the blockchain \label{rdvline:35}
			\State 
			Insert deposited coins of voters whose votes was 0 in list of blocked coins 
			meaning that they are unacceptable to be sent for next transactions. 
			
			\Else \label{rdvline:37}
			\State 
			Insert deposited coins of voters whose votes was 1 in list of blocked coins 
			meaning that they are unacceptable to be sent for next transactions.
			
			\EndIf \label{rdvline:39}
			
			\State Cleans NotParticipatedSinceDeltaList  \label{rdvline:40}
			\EndIf \label{rdvline:41}
			\EndWhile \label{rdvline:42}
		\end{algorithmic}
	\end{algorithm*}
	\thispagestyle{empty}
	\clearpage} 

\subsection{Correctness of RDV Consensus}
\label{Correctness}

We divide the nodes into two sets: ordinary and registered. Then we define a set of registered nodes as follows: \newline

$ RNset  = {rn_{1}, rn_{2}, \dots, rn_{n}} $ \newline

Where, \textit{rn} is a register node and \textit{RNset} is set of all current registered nodes. \\

$ state_{1} $: If all registered nodes participate in voting process within $ \Delta $ time unit and all of votes are equal, then we achieve consensus, otherwise we define $ state_{2} $ as follows:

$ state_{2} $: All registered nodes participate in voting process within $ \Delta $ time unit, however all of votes are not equal. Then, the majority vote value (1 or 0) will be considered as dominant vote. Additionally, voters who their vote is not equal to this dominant vote will lose a part of their deposited coins as a penalty and the rest of voters receive a CTR reward to be motivated (see CTR in section \ref{Definitions}).

$ state_{3} $: If one registered node doesn't participate in voting process within $ \Delta $ time unit, then, this node will be removed from $ RNset $ 
for $ \Pi $ time unit as a penalty (see section \ref{RDV_Flowchart}). As a result, (n-1) \textit{rn} will achieve a consensus. 

$ state_{4} $: If two registered nodes do not participate in voting process within $ \Delta $ time unit, then, these two \textit{rn} will be removed from $ RNset $ 
for $ \Pi $ time unit as a penalty. As a result, (n-2) \textit{rn} will achieve a consensus. 

We continue this process till $ state_{n-1} $ as follows:

$ state_{n-1} $: (n-1) registered nodes do not participate in voting process within $ \Delta $ time unit. Then, (n-1) \textit{rn} will be removed from $ RNset $ 
for $ \Pi $ time unit as a penalty. As a result, one \textit{rn} determines the result. 

And finally $ state_{n} $:

$ state_{n} $: All of registered nodes do not participate in voting process within $ \Delta $ time unit. Then, all of them will be removed from $ RNset $ and after joining new registered nodes we will have another new set of registered nodes ($ RNset $). Then, we go to $ state_{1} $. Thus eventually we achieve a consensus.


\paragraph{What happens when a user attempts to cheat and presents an old time-stamp to increase the Priority Point of his transaction?}  Since the information propagation is calculable (e.g. \textit{m} time unit) so, if an adversary node intends to forge the time-stamp to e.g. \textit{(m + 10)} time unit ago, then the question of honest nodes is why this transaction has not been broadcast 10 time units ago (i.e. immediately after doing transaction)? Thus, there is some issues in this transaction. Moreover, rational node is able to forge the time-stamp, if the other side of transaction (i.e. receiver or sender) is adversary as well. \par

\paragraph{How to bootstrap such a system?} The initial deposit has a negative value. It means that initially a voter deposits \textit{-d} coins. Then, in case of winning, they get some rewards and so they have enough coins for the next time. And if they have to pay some penalty, the first time they receive some coins (e.g. \textit{r} coins), then they will have \textit{(r - d)} coins.\par

\begin{table}[h!]
	\centering
	\begin{tabular}{| c | c | c | c |} 
		\hline
		transaction  &  coin  &  sender address  &  Priority Point  \\ [0.5ex] 
		\hline\hline
		$ \mathrm{tx_{i}\rightarrow timestamp\hspace{0.10cm}, CTR} $ & $ \mathrm{coin_{i}} $ & $ \mathrm{address_{i}} $ & $ \mathrm{max} $ \\ 
		\hline
		$ \mathrm{tx_{j}\rightarrow timestamp\hspace{0.10cm}, CTR} $ & $ \mathrm{coin_{j}}  $ & $ \mathrm{address_{j}} $ & ... \\
		\hline
		$ \mathrm{tx_{k}\rightarrow timestamp\hspace{0.10cm}, CTR} $ & $ \mathrm{coin_{k}} $ & $ \mathrm{address_{k}} $ & ... \\
		\hline
		... & ... & ... & ... \\
		\hline
		$ \mathrm{tx_{m}\rightarrow timestamp\hspace{0.10cm}, CTR} $ & $ \mathrm{coin_{m}} $ & $ \mathrm{address_{m}} $ & $ \mathrm{min} $ \\ 
		\hline 
	\end{tabular}
	\caption{The Priority Point Table.}
	\label{TablePeriority}
\end{table} 

\subsection{Preventing Double-Spending Attack by RDV}
\label{Double-Spending}

In fact, a rational user performs a race attack to be winner in double-spending \cite{Xiaoqi}. This attack is easier to perform in proof-of-work based blockchain compared  with other type of blockchains \cite{Xiaoqi} and the reason is latency of this type of consensus mechanism \cite{misbehavior_in_bitcoin} and thus the attacker has enough time for double-spending. \par 

Authors in \cite{Double-spending} analysed this type of attack in fast payment transactions in Bitcoin. They proposed an attack in which if three conditions are met, then rational node can receive the expected item (i.e. the vendor’s service) without spending any coin: 

\begin{enumerate}
	\item If $ tx_{v}  $(transaction sent to vendor) is added to the vendor’s wallet.
	\item $ tx_{a} $ (transaction sent to a colluding mining pool) is inserted into blockchain.
	\item rational node receives expected item from vendor before double-spending is detected. In such a situation, rational user without spending any coins receives his item from a vendor.
\end{enumerate}

\begin{figure*}
	\centering
	\makebox[\textwidth][c]{\includegraphics[width=1\textwidth]{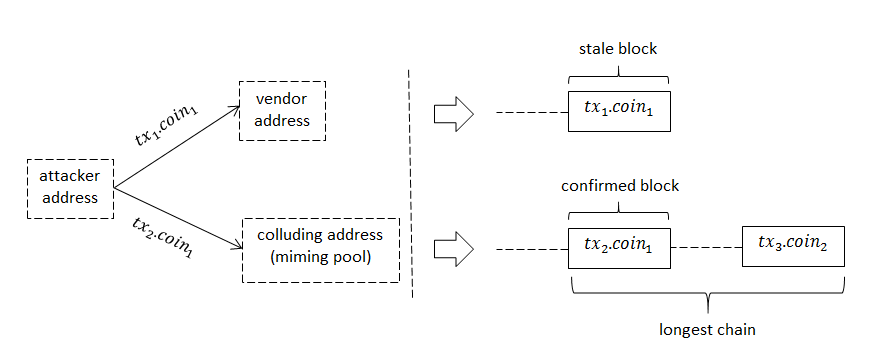}}
	\caption{a double-spending attack.}
	\label{double_spending_fig}
\end{figure*}

In RDV, double-spending is impossible or very difficult. Because every transaction has need for vote of all current registered nodes, where the list of voters is updated periodically by $\Delta$ time unit (see section \ref{Definitions}).  As a result, as long as  majority of voters are honest, every coin that is spent more than one time is recognised by the priority table, Table \ref{TablePeriority}, and equation \ref{eq-Double-Spending}.    

\begin{equation}
i\!f\hspace{0.10cm} tx_{i}\rightarrow[coin + address] == tx_{j}\rightarrow[coin + address]\hspace{0.10cm} then
\label{eq-Double-Spending}
\end{equation}
\begin{center}
	\textit{a double-spending occurred}
\end{center} 
\vspace{5mm}

where \textit{tx} is transaction, \textit{address} is sender's address and '+' is concatenation operation.

\subsection{Preventing Blockchain Fork and Block-withholding by RDV}

Block-withholding attack was introduced as "Selfish mining attack"  in \cite{Eyal} and also as "Block Discarding Attack" in \cite{Bahack}. This attack relies on "block concealing" and revealing only at a specific time selected by selfish miners or selfish mining pool.
According to \cite{Eyal}, these selfish miners can earn revenues superior to a fair situation \cite{Luu}.  
That is, the main purpose of block-withholding by selfish mining pool is achieving more rewards in comparison with its hashing power in the network. Thus, selfish mining pool's reward oversteps its mining power in the network and it can increase its expected mining reward. This attack leads to blockchain fork. 

Although, some solutions are proposed to prevent this attacks \cite{ZB1,Heilman}, however, in RDV, unlike Bitcoin, the new state of blockchain is not broadcast, but also list of voters signed by all voters (\textit{voteRbox}) and list of signed votes (\textit{voteBox}) are broadcast to the network. So, there is no possibility for block-withholding and forking blockchain intentionally by adversary. \\

On the other hand, since proof-of-work is a Poisson process, two blocks may be discovered by two mining pools, almost at the same time. We removed the Poisson nature of proof-of-work that causes accidental fork in Bitcoin. Instead, the next transaction for being participated in voting process is selected by Priority Point table, i.e. Table \ref{TablePeriority}. Transactions in this table are selected \textbf{sequentially} according to their Priority Point.

\subsection{Removing Some Parameters And Criteria}

In Pow consensus mechanism, miners try to adjust their strategy to participate in a "speed game" (i.e. solving PoW puzzle) in which they must produce a new block with most difficulty (i.e. longest chain) as soon as possible. The "longest chain" parameter leads to a motivation for "rational" miners to perform a block-withholding attack \cite{Eyal} and forking the blockchain. 

In RDV, we remove these parameters and benchmarks to avoid these existing problems of the PoW based blockchains, such that in RDV, there is no mining process and as a result the difficulty of a chain is not a parameter by which a new block can be judged or decided.

\subsection{Immutability of Transactions History}

In RDV consensus, transactions history is immutable because of following reasons. \\

First of all, note that according to the RDV algorithm (Figure \ref{fig_2}), unlike Bitcoin, the new state of blockchain is not broadcast, but also list of voters signed by all voters (voteRbox) and list of signed votes (voteBoxes) are broadcast to the network. \\

\begin{figure*}
	\includegraphics[scale=0.31]{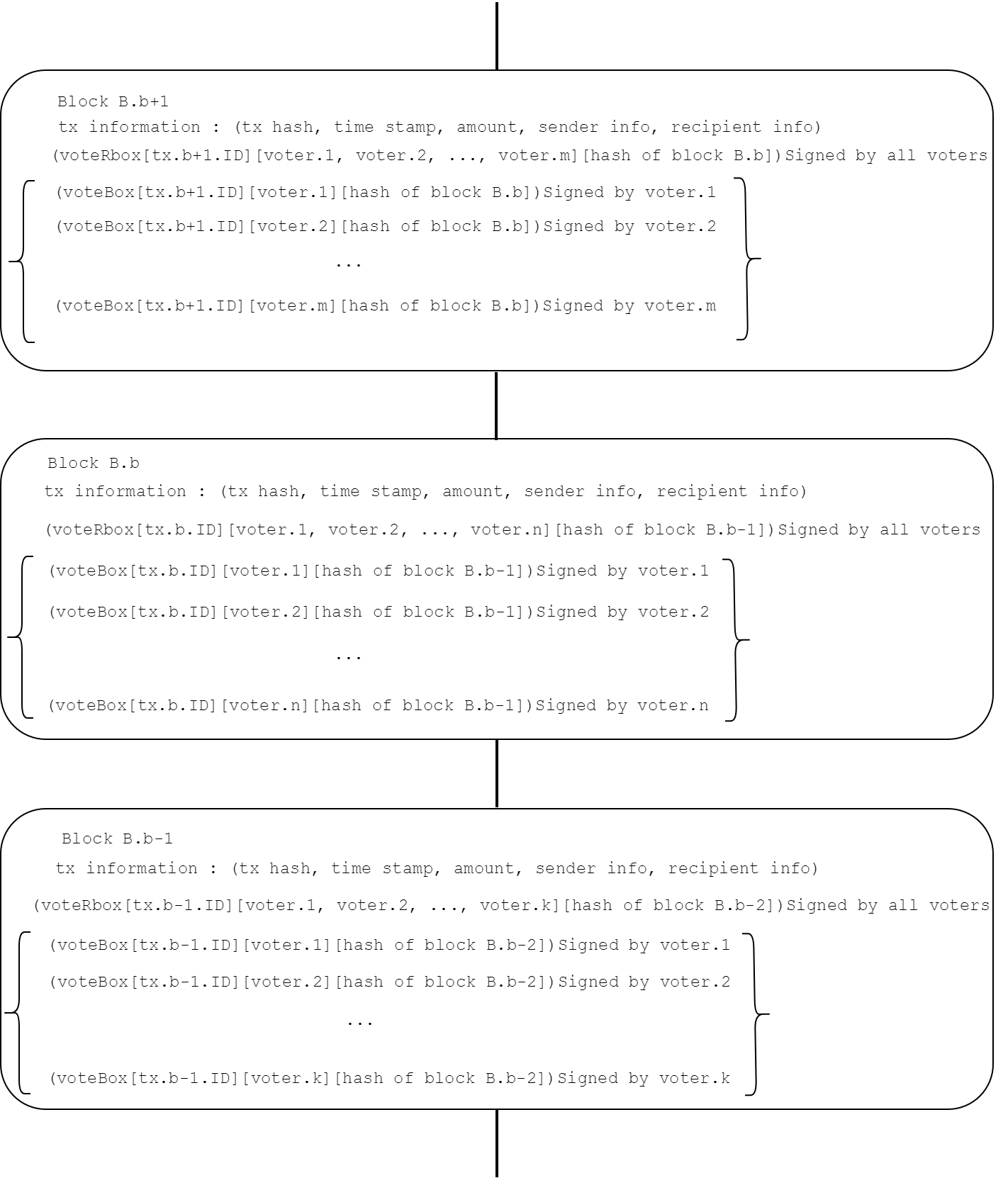} 
	\caption{Block structure in RDV consensus.}
	\label{fig_4}
\end{figure*}

On the other hand, each block $ B_{b} $ includes \textit{voteRbox} and \textit{voteBoxes} as follows (see also Figure \ref{fig_4}): \\

\begin{math}
\label{eq-voteBox}
\mathrm{(voteBox[tx.ID][voter ID][Hash(B_{b - 1})])_{Signed \hspace{0.10cm} by \hspace{0.10cm} voter \hspace{0.10cm} private \hspace{0.10cm} key}}
\end{math}

\vspace{5mm} 

\begin{math}
\label{eq-voteRbox}
\mathrm{(voteRbox[tx.ID][list \hspace{0.10cm} of \hspace{0.10cm} voters][Hash(B_{b - 1})])_{Signed \hspace{0.10cm} by \hspace{0.10cm} all \hspace{0.10cm} voters}}
\end{math}

\vspace{5mm} 

Where, $ \mathrm{Hash(B_{b - 1})} $ contains hash of block $ \mathrm{B_{b - 1}} $. \\

So, if adversaries make any changes in block $ B_{b - 1} $ then all blocks after $ B_{b - 1} $ including block $ B_{b} $ will become \textbf{invalid} since it includes hash of block $ B_{b - 1} $ signed by all voters of \textit{voteRbox}. \\

On the other side, because \textit{voteRbox} is signed by all voters who are included in it, so adversaries need to forge signatures of other voters who are not included in their cartel to make any changes in the list of voters and values of the votes. 

\begin{figure*}
	\includegraphics[scale=0.55]{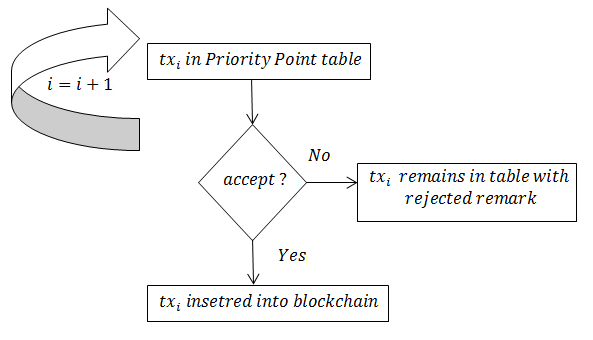} 
	\caption{A voter can check multiple transactions validation in parallel.}
	\label{fig_3}
\end{figure*} 

\subsection{Increasing Transactions Confirmation Throughput}

While in PoW a miner needs very fast processors to do mining process, in RDV, a voter can participate in voting process by a very ordinary computer. On the other hand, while mining process to solve PoW puzzle takes significant time (since because of security issues we need to keep difficulty of PoW high enough), in RDV there is no mining process and it causes increasing transactions confirmation throughput. As Figures \ref{fig_3} and \ref{fig_2} show, a voter can check multiple transactions validation in parallel. \\

\subsection{Block Structure in RDV Algorithm}

In this section, we show the structure of blocks in RDV consensus mechanism in which each block consists of the list of all voters for that transaction signed by all of them (i.e. \textit{voteRbox}) along with their signed votes (i.e. \textit{voteBox}). Both of these lists include hash of previous block. This feature causes that if adversaries make any changes in block $ B_{b - 1} $ then all blocks after $ B_{b - 1} $ including block $ B_{b} $ will become \textbf{invalid}, because it includes hash of block $ B_{b - 1} $ that is signed by all voters of \textit{voteRbox}. On the other side, because \textit{voteRbox} is signed by all voters who are included in it, so adversaries need to forge signatures of other voters who are not included in their cartel to make any changes in the list of voters and values of the votes.    

\subsection{Main Problems of Proof-of-Work}

Here we explain our motivation to propose an alternative to PoW. We mention most important vulnerabilities and security problems of proof-of-work as follows. At the same time, we explain how RDV improves these weaknesses and resists these problems better than proof-of-work. 

\begin{itemize}

	\item \textbf{Energy Consumption:}
	Bitcoin uses significant amount of energy because of nature of PoW mechanism. This can lead to a considerable problem in long term. 
	The expected electricity for Bitcoin mining has been debated over the past few years. The mining process makes Bitcoin very \textbf{energy-hungry} system where it needs a significant amount of hash computations. The main resource of these process is electricity. It has been estimated the Bitcoin network currently consumes 2.55 GW of electricity at the least and potentially 7.67 GW in the future. These amounts are comparable to countries like Ireland (3.1 GW) and Austria (8.2 GW) \cite{Growing_Energy_Problem}. \\
	
	\item \textbf{Monopoly Problem and Decentralization of the Network:} If a miner can take majority of transactions verification resources (i.e. mining calculation power), then this miner is able to impose the conditions on the rest of the network. This problem is known as "monopoly" and this miner is called as "monopolist". The monopolist can be malevolent or benevolent. The malevolent monopolist performs malicious strategies such as double-spending (race attack) or DoS attack. If monopolist entity can keep this situation for a long term, then cryptocurrency reputation will be undermined. Even in case of benevolent monopolist, cryptocurrency and network reputation is hyper-dependent on this entity's decision. At this point, decentralization of the network is collapsed. Trying for being monopolist increases progressively over time, because of the concept of the \textbf{Tragedy of the Commons}. 
	Decentralization of the Bitcoin network is currently under a risk. Since finding the correct answer of PoW and as a result block generation is (and must be) very difficult (because of security issues), so only mining pools with significant hashing power are able to determine the fate of transactions. On the other side, if a mining pool achieves more than 51 percentage of total hashing power of the network, then according to 51\% attack, this mining pool is able to control the network, where \textbf{the network cannot control the cost of requirements for mining operation (such as the cost of GPUs, ASIC etc) and this means that the network is not able to control the cost of attack for an adversary}. An organization already achieved more than 51\% hashing power of the network \cite{miner-dilemma} and so it may happen again in the future. And in this case, "we will be forced to trust" this mining pool with such a hashing power. At this point, we can say that "decentralization of the network is collapsed". This might mean \textbf{we have nothing}, when the main purpose of using blockchain technology is maintaining the system with a real "decentralized" approach. 
	
	On the other hand, in Bitcoin consensus mechanism, because of difficulty in solving the PoW cryptographic puzzle, there is a significant latency in block generation.This causes inserting several transactions in one block that affects negatively decentralization of the network. However, in RDV, each block consists of only one transaction, so this can lead to increasing network decentralization. \\
	
	Despite belief that PoW has an acceptable scalability as a  lottery-based algorithm due to no need to exchange messages \cite{Hyperledger_Consensus} ; however, if we define PoW as one-cpu-one-vote, then with growing the network hashing power of the network will be increased and as a result, we need to increase difficulty of PoW that causes participating in transactions validation (aka mining process) would be more difficult for miners who do not possess enough fast processors and this situation continues till for participating in mining process must be joined to a large mining pool. This process, in a long time, causes the blockchain would be controlled by some large and limited mining pools. This eventually affects negatively decentralization of the blockchain.
	
	
	
	\item \textbf{Latency:} The PoW is based on a cryptographic puzzle that is difficult to solve but easy to verify. The security of PoW is relied on the difficulty of PoW, meaning that if we decrease the difficulty to accelerate the transaction validation, then it affects negatively the security and consistency. Apart from delay in transactions validation, this latency causes also vulnerability against double-spending \cite{misbehavior_in_bitcoin}. While mining process to solve PoW puzzle takes significant time, in RDV there is no mining process and it causes increasing transactions confirmation throughput. \\
	
	
\end{itemize}

\subsection{Comparing with Proof-of-Stake}

Consensus may be designed in different approaches: The first category is lottery-based algorithms such as Proof of Work (PoW) and Proof-of-Stake (PoS) in which the winner of the lottery proposes a block and transmits it to the rest of the network for validation \cite{Hyperledger_Consensus}. On the other side, we have voting-based approaches such as Redundant Byzantine Fault Tolerance (RBFT) \cite{RBFT}, Practical Byzantine fault tolerance (PBFT) \cite{PBFT} and Paxos \cite{Paxos}. The lottery-based consensuses may lead to forking when two winners propose a block almost at the same time. Each fork must be fixed that causes a longer time to finality \cite{Hyperledger_Consensus}. The main difference between RDV and PoS is that while PoS is a lottery based algorithm, RDV is based on distributed voting process and as we explained before there is no fork in RDV based blockchain. So, one of the main security problems of PoS i.e. “nothing at stake problem” that is occurred because PoS is lottery based consensus and there is possibility of forking blockchain, in RDV there is not such type of attacks since as we mentioned RDV is a voting based algorithm and there is no fork in RDV based blockchain. We will explain in more details the PoS and related problems such as “nothing at stake problem” as follows.

\subsubsection{nothing at stake problem}

The first versions of proof-of-stake did not need a security deposit such that the users only required owning tokens in order to be permitted for being a validator, such that having tokens in the wallet was as the user’s stake. In the case these validators attacked the network, it did not affect their coins (as their stake). However, in the upgraded version the stake referred to as the deposited tokens that validators had to send before they were permitted to propose blocks. 
The idea behind PoS was that stakeholders with more coins are less likely to destroy the system since if the blockchain was effectively attacked, then the value of stakeholders tokens was probable to considerably drop.
The nothing at stake problem is based on the assumption that, every stakeholder will build on every fork whenever a fork occurs. This assumption is based on two following reasons:

\begin{itemize}
	
	\item Contrary to PoW, it costs nothing for a stakeholder to confirm transactions on several forks, because they no longer require solving PoW puzzle to create a block. \\
	
	\item Stakeholders likely build on every fork since if they expand several chains, then they’ll gain more fees on whichever fork winds up winning. This behaviour affects negatively consensus and may lead to make the system more vulnerable to double-spending. However, in PoW based blockchains, the incentive to mine on several chains at the same time causes miners split their calculation power among several chains such that it does not enhance their chance to be winner. \\
	
\end{itemize}

As a result, Ethereum (Casper version) intends to prevent this problem by adding a penalty for validators, meaning that losing a part or all of their deposited tokens.
However, as we explained, the functionality of RDV is essentially different since it is a voting based consensus algorithm and there not possibility of forking RDV based blockchain.

\section{Discussion and Future Works: Enhancing Security of the System by Combination of Interior and Exterior Resources}

We proposed a new consensus mechanism, RDV, as a more decentralized alternative to PoW because of its major vulnerabilities, security problems, energy consumption, latency etc. We proved the correctness of RDV consensus. We showed that RDV is more democratic, fairer and more decentralized than PoW and it can resist major problems such as double-spending, forking blockchain, immutability of transactions history and transactions confirmation throughput better than PoW by achieving a more decentralized system. RDV algorithm has an \textbf{incentive-punitive} mechanism since for achieving an ideal crypto-currency, we need to design an appropriate incentive-punitive system such that according to the Nash equilibrium, diverging from the protocol does not lead to a net profit for the adversary \cite{Kroll}. \\
RDV is based on distributed voting process and since in RDV algorithm, there is no mining process, so it is appropriate for low-level energy devices and Internet of Things (IoT). \\

In general, security of a system must not be depended on only \textit{exterior} resources. For example, in Bitcoin, if a miner has access to a cheap or free electricity resources, then the network faces significant security risks. Since the cost of necessary electricity for mining process is \textbf{not controllable} via \textbf{inside} of the system, so we are not able to control security of the system as it should be. On the other hand, if miners have access to free or cheap electricity resources, they do not spend an adequate penalty for their malicious behaviour such as forking blockchain. In other words, forking blockchain has no considerable cost for the adversary. Thus, the security parameters \textbf{must be controllable} via the \textbf{inside} of the system. \\

However, exterior resources can be very useful as a "complementary" parameter, meaning that exterior resources can be employed as a complementary, but the main resources for ensuring security of the system must be chosen from inside of the system (ex. coins in a crypto-currency network). Consequently, in this way, we intend to extend the RDV algorithm by adding an external cost for participating in transactions confirmation. Regarding the fact that relying on mining process is very energy-hungry mechanism, so we focus on using approaches such as Proof-of-Space-Time (PoST) \cite{PoST}, such that every node before vote for a transaction has to prive spending a space-time resource, meaning that the storage over a period of time (based on the approach described in \cite{PoST}). We try to keep supporting fee-free transactions and at the same time avoiding spam transactions by adding an external cost for adversary, but contrary to PoW that is CPU / processor based, we focus to use memory based approached. We also keep voting based approach (instead of lottery based) since they have voting-based algorithms are advantageous in that they provide low latency finality and better avoiding blockchain fork \cite{Hyperledger_Consensus}. So, we use memory as an complementary external cost to prevent and control the attacks more efficiency by \textbf{imposing more cost on the adversary}. The approach based on spending "space-time" is more flexible by having two parameters: spending (1) storage and (2) time, such that we can adjust its parameters more efficiency for the low level energy devices by ex. relying more on the time parameter than the memory depending on the equipments. \\

So, this approach will increase the cost of attacks for an adversary by using both internal cost (i.e. the coins pledged as collateral by voter at time of registration) and external cost (i.e. spending a space-time resource), whereas PoW relies only on external cost (i.e. processor and electricity cost for mining process) while the cost of an external resource \textbf{is not controllable} by the network. 

\section{Other Blockchain Consensuses And Their Vulnerabilities}

\begin{itemize}
	
	
	\item Practical Byzantine Fault Tolerance: PBFT is a replication algorithm that is able to tolerate Byzantine faults \cite{PBFT} up to 1/3 malicious byzantine replicas. One of blockchain platforms that employs PBFT is Hyperledger Fabric \cite{hyperledger_fabric}. In this approach a new block is created in a round in which a primary will be selected based on several rules. Then, it will be responsible for ordering transactions. The process is divided into three phases as follows: pre-prepared, prepared and commit, such that each node is permitted to enter the next phase if it has gotten 2/3 votes of all nodes. As a result, the nodes in PBFT must be known to the network. One of the PBFT problems is its scalability, because every node must send messages to every other node, such that for n nodes, the number of required messages are n(n-1) (with complexity of O($ n^{2} $)). So, it is not scalable to large networks and as a result, it is used in a permissioned blockchain where number of permitted validators is limited.
	
	\item Delegated-Proof-of-Stake: DPoS \cite{dpos_bitshare} is a specific type of PoS in which stakeholders choose their delegates to validate transactions and create new block such that the number of nodes who validate transactions considerably is decreased and so the new block can be confirmed faster. Also, block size and block interval time can be adjusted by selected delegates. Those delegates might be deselected by stakeholders. The main security problems of PoS remain with DPoS.
	
	\item Ripple: This consensus approach \cite{Ripple} uses collectively-trusted sub-networks within the larger network. It divides the nodes into two types: server and client. While the servers participate in consensus, the clients are only permitted to send funds. every server owns a unique list, named UNL (Unique Node List). At time of deciding if a transaction must be inserted into the ledger, the server sends a query to its UNL such that if the received agreements have reached 80\%, transaction will be put into the ledger. In this protocol, till the number of faulty nodes in UNL is less than 20\%, the ledger will be accepted as a correct one. 
	
	
	The ripple consensus algorithm (RPCA) is a round based mechanism in which in every round:
	
	\begin{enumerate}
		
		\item Servers take valid transactions which have not been previously applied. Then, they publish them in a form a list, "candidate set".
		
		\item Then,every server combine all servers candidate sets on his Unique Node List. Then, they vote on accuracy of transactions.
		
		\item Transactions which get more than a minimum percentage of "positive votes", are authorized to pass to next round. The rest of transactions will be discarded, or inserted in candidate set to be waited for the next ledger.
		
		\item Last round needs for a minimum percentage of 80\% of a server's Unique Node List agreeing on a transaction. 
		
		\item The eligible transactions are inserted to the ledger. the ledger then is closed to become a new Last Closed Ledger.
		
	\end{enumerate}
	
	\paragraph{Ripple Protocol Components:}
	
	The Ripple protocol consists of following components:
	
	\begin{enumerate}
		
		\item Server: There are two types of Ripple software: Ripple Server software which runs by an entity, called as Server that participates in consensus protocol. And Ripple Client software that only permits a user to send and receive the funds.
		
		\item Ledger: It holds the amount of currency in every account in the network and is updated periodically with transactions after via a consensus mechanism.
		
		\item Last Closed Ledger: It is the most recent ledger which has been approved after a consensus. It represents the current state of the network.
		
		\item Open Ledger: Every node holds its local open ledger. Transactions in the open ledger are not considered till they have been approved via a consensus mechanism when the open ledger becomes last closed ledger.
		
		\item Unique Node List (UNL): Every server \textit{s} holds a UNL which consists of a set of other servers that are queried by \textit{s} at the time of determining consensus. 
		
		\item Proposer: Every server is able to broadcast the transactions to be included in consensus mechanism. They also try to include a valid transaction at the time a new consensus round.   
		
	\end{enumerate}
	
	\paragraph{Mining in Ripple Protocol:}
	
	Ripple is based on a blockchain-similar mechanism. However, unlike Bitcoin network, it does not need for an energy-consumer mining process and it is only based on a consensus mechanism. This technology is employed by large organizations e.g. banks. \par 
	
	The main purpose of this idea is to permit financial institutions to transfer any type of asset (e.g. currencies, gold, etc).  \par 
	
	\item Tendermint: This approach \cite{Tendermint} is a Byzantine and round based consensus algorithm in which a new block will be created in a round, such that in a round a proposer is selected to broadcast a new block that is not yet confirmed. In the first step, validators decide about broadcasting a "Prevote" for a proposed block. Then, If a node receives more than 2/3 of
	"Prevotes" for the proposed block, it will broadcast a "Precommit" for proposed block. Then, If the node receives over 2/3 of "Precommits", then node confirms the block and broadcasts a commit confirmation for proposed block. Eventually, if the
	node receives 2/3 of the commits, then it accepts the new block. Contrary to PBFT, the nodes need to lock their coins to be a validators.
	
\end{itemize}

\end{document}